\documentclass[12pt]{article}

\usepackage{amsmath}
\usepackage{braket}
\usepackage{amssymb}
\usepackage{amsfonts}
\usepackage{latexsym}
\usepackage{graphicx}
\usepackage{latexsym}
\usepackage{color}
\usepackage{cite}
\def\ln{\,\mbox{ln}\,}

\def\al{\alpha}
\def\be{\beta}
\def\ga{\gamma}
\def\Ga{\Gamma}
\def\de{\delta}

\def\ka{\kappa}

\def\ph{\varphi}

\def\om{\omega}

\def\pa{\partial}

\def\beq{\begin{eqnarray}}
\def\eeq{\end{eqnarray}}

\def\lap{\Delta}

\begin{document}

%%%%%%%%%%%%%%%%%%%%%%%%%%%%%
\begin{center}

{\Large%\bf
Effective delta sources and regularity in \\
\vskip 2mm
higher-derivative and ghost-free gravity
}
\vskip 8mm

{\large
Breno L. Giacchini$^{a}$
\ \ and \ \
Tib\'{e}rio de Paula Netto$^{b}$}

\end{center}
%%%%%%%%%%%%%%%%%%%%%%%%%%%%%
\vskip 1mm

%%%%%%%%%%%%%%%%%%%%%%%%%%%%%
\begin{center}
{\sl
(a) \ Centro Brasileiro de Pesquisas F\'{\i}sicas
\\
Rua Dr. Xavier Sigaud 150, Urca, 22290-180, Rio de Janeiro, RJ, Brazil
\vskip 3mm

(b) \ Departamento de F\'{\i}sica, \ ICE, \ Universidade Federal de Juiz de Fora, 
\\
Campus Universit\'{a}rio, 36036-330 Juiz de Fora, \ MG, \ Brazil
\vskip 3mm
}
\vskip 2mm\vskip 2mm

{\sl E-mails:
\ \
breno@cbpf.br,
 \
tiberiop@fisica.ufjf.br}

\end{center}
%%%%%%%%%%%%%%%%%%%%%%%%%%%%%
\vskip 6mm

\begin{quotation}
\noindent
\textbf{Abstract.}
It is shown that polynomial gravity theories with more than four derivatives in each scalar and tensor sectors have a regular weak-field limit, without curvature singularities. This is achieved by proving that  in these models the effect of the higher derivatives can be regarded as a complete regularization of the delta-source. We also show how this result implies that a wide class of non-local ghost-free gravities has a regular Newtonian limit too, and discuss the applicability of this approach to the case of weakly non-local models.
\vskip 3mm

%{\it MSC:} \
%53B50,  %%%	    Applications to physics
%83D05,  %%%    Relativistic gravitational theories other than
              %%%     Einstein's,   including asymmetric field theories
%81T20	  %%%     Quantum field theory on curved space backgrounds
%%%%%%%%%%%%%%%%%%%%%%%%%%%%%%%%%%%%%
%\vskip 2mm

PACS: $\,$
%04.62.+v,	 %%%%%% Quantum fields in curved spacetime
04.20.-q,     %%%%%% Classical general relativity
04.50.Kd 	 %%%%%% Modified theories of gravity
\vskip 2mm

\end{quotation}

\section{Introduction}
\label{Sec1}

The inclusion of higher-derivative curvature-squared terms into the Einstein-Hilbert action has been proved to smooth the quantum and classical divergences which usually stem in the ultraviolet (UV) regime. On the one hand, since the 1970s it is known that general relativity (GR) is not perturbatively renormalizable~\cite{tHooftVeltman74,GoroffSagnotti86}, while its counterpart with four derivatives is~\cite{Stelle77} --- and local models with at least six derivatives become superrenormalizable~\cite{AsoreyLopezShapiro}. On the other hand, from the classical perspective, whereas Newton's potential diverges at the origin, Stelle's fourth-order gravity has a finite non-relativistic potential~\cite{Stelle77} (but still has curvature singularities when coupled to a $\delta$-function source~\cite{Stelle78,Stelle15PRL,Stelle15PRD}). Recently it was shown that polynomial-derivative theories
of order higher than four
have a regular Newtonian limit. Indeed, in addition to having a finite potential~\cite{Newton-MNS,Newton-BLG} they also have regular curvature invariants~\cite{BreTib1}.

In this work we discuss how this increasing regularity of the Newtonian-limit solutions can be viewed as a regularization of the $\delta$-source, as one includes higher derivatives. Such analogy is frequent in the framework of non-local gravity theories and can be tracked back to~\cite{Tseytlin95} (see also~\cite{Modesto12,Zhang14,Li15,Myung17,Buoninfante:2018b,Jens} and references therein for more recent considerations). Even though this analogy also holds in the case of local higher-derivative gravity (HDG), it is seldom discussed in the literature. In fact, to our knowledge the only examples of this kind are~\cite{Jens,Modesto-LWBH}, where particular cases of Lee-Wick gravity were analysed. Therefore, it is instructive to extend considerations to general higher-derivative polynomial theories, including the cases in which the propagator has complex and/or degenerate poles.

Most of our discussion concerns the gravity model defined by the action\footnote{We use the same sign conventions as~\cite{BreTib1}. Also, we set $c\,=\,\hbar\,=\,1$.}
\beq
\label{act}
S_\text{grav} = \frac{1}{4 \ka} \int d^4 x \sqrt{-g} \,
\Big\{ 2  R 
+ R_{\mu\nu} \, F_1 (\Box)  \, R^{\mu\nu} 
 + \, R \, F_2(\Box) \, R
\Big\}\,,
\eeq  
where $\ka = 8 \pi G$ and $F_1$ and $F_2$ are polynomial functions of the d'Alembert operator, not necessarily of the same degree. We shall also consider non-local higher-derivative theories~\cite{Modesto12,Krasnikov,Kuzmin,Tomboulis,BMS06,BGKM12}, in which $F_1$ and/or $F_2$ are non-polynomial functions
with improved UV behaviour\footnote{That is, we shall consider non-local gravity models which are extensions of GR in the UV-limit, which means that for large momentum the propagator decays faster than in GR. Specifically, we require that $f_0(z)$ and $f_2(z)$ (defined in~\eqref{f0} and~\eqref{f2}) are constant or diverge at least linearly as $z \longrightarrow \infty$, and that $f_s(0) = 1$. Owed to this improved behaviour in the UV, sometimes these models are called non-local HDG. The situation is quite different from non-local IR modifications of GR, such as those defined by form factors of the type $F_i \propto \Box^{-1}$ and $F_i \propto \Box^{-2}$~\cite{DesWood,Maggiore1,Maggiore2}, or the logarithmic ones, $F_i \propto \ln \Box$, which  come from the integration of quantum matter fields in curved space-time~\cite{apco1,apco2,apco3,Netto:2016imv,Franchino-Vinas:2018gzr}.}.
We recall that at the linear regime a term of the type $R_{\mu\nu\al\be} F_3 (\Box) R^{\mu\nu\al\be}$ can be recast as a combination of Ricci- and $R$-squared terms (see, e.g.,~\cite{AsoreyLopezShapiro}). Hence, to our purposes the action~\eqref{act} is the most general one with higher derivatives.

The propagator associated to the model~\eqref{act} can be written in the momentum-space representation as
\beq
\label{prop}
D_{\mu\nu,\alpha\beta} (k)
=
 \frac{P^{(2)}_{\mu\nu,\alpha\beta} }{k^2 f_2(-k^2)}
- \frac{P^{(0-s)}_{\mu\nu,\alpha\beta} }{2k^2
f_0(-k^2)
} \,,
\eeq
where $P^{(2)}_{\mu\nu,\al\be}$ and $P^{(0-s)}_{\mu\nu,\alpha\beta}$ are the spin-2 and spin-0
projectors (see, e.g., \cite{book-Sh}); the other terms which are gauge-dependent have been omitted. The functions $f_0$ and $f_2$ are related to $F_1$ and $F_2$ through
\beq
f_0(\Box) &=& 1  - F_1(\Box)\Box - 3F_2(\Box) \Box  \,, 
\label{f0}
\\
f_2(\Box) &=& 1 + \frac12\, F_1(\Box) \, \Box  \, .
\label{f2}
\eeq
Therefore, the roots $-m_{(s)i}^2$ of the equation $f_s(-k^2) = 0$ define the massive poles of the propagator. The index $s=0,2$ indicates the spin of the particle, while $i$ enumerates the particles of the same spin. If the functions $f_s$ are polynomials, then many of these massive excitations correspond to ghost modes~\cite{AsoreyLopezShapiro}.

Two proposals for dealing with the ghosts (or avoiding them) in the framework of  HDG have been the subject of intensive investigation in the recent years. The first possibility we mention is accomplished by requiring that the massive poles in the propagator which are associated to ghost modes are complex. Then, from the quantum gravity perspective, the theory has a unitary S-matrix in the Lee-Wick sense~\cite{ModestoShapiro16,Modesto16}. Different aspects of such Lee-Wick gravity theories have been considered in, e.g.,~\cite{Newton-BLG,Modesto-LWBH,Jens,ModestoShapiro16,Modesto16,Seesaw,ABS-large,Lens-LWBH}. The second possibility consists in avoiding ghosts by choosing functions $F_i$ such that $f_s$ is the exponential of an entire function~\cite{Modesto12,Krasnikov,Kuzmin,Tomboulis,BMS06,BGKM12}. Consequently, the equation $f_s(-k^2)=0$ has no roots in the complex plane, and no other excitations besides the graviton are present. The most simple example of non-local ghost-free gravity is to set $f_s$ to be the exponential of a monomial. Other interesting choices for $f_s$ are the so-called weakly non-local form factors, which have the same behaviour as a polynomial in the UV domain~\cite{Modesto12,Kuzmin,Tomboulis}.

In the following sections we investigate the Newtonian limit of polynomial HDG coupled to a $\delta$-source and explicitly show that the occurrence of regular solutions can be viewed as a regularization of an effective source. The considerations are quite general, comprising the models with complex poles (Lee-Wick gravity) and/or higher-order poles in the propagator. In section~\ref{Sec5} we discuss the case of  non-local ghost-free HDG. It is shown that for a wide class of theories the corresponding effective sources can be obtained as the uniform limit of a sequence of sources associated to polynomial HDG, and hence the aforementioned regularity properties also hold.

\section{Newtonian limit}
\label{Sec2}

In the static weak-field regime we consider the metric to be a fluctuation around the flat Minkowski space-time, $g_{\mu\nu} = \eta_{\mu\nu} + h_{\mu\nu}$, and 
restrict considerations to the linearised equations of motion. Applying the variational principle to the action~\eqref{act} supplemented by a matter action, one gets the equations for the field $h_{\mu\nu}$:
\beq
\label{lieq}
&&
f_2 (\Box) \, (\Box h_{\mu\nu} - \pa_\rho \pa_\mu h^\rho_\nu - \pa_\rho 
\pa_\nu h^\rho_\mu)
+ \frac{1}{3} \left[ 2 f_0(\Box)+  f_2(\Box) \right]  \, (\eta_{\mu\nu} \pa_\rho \pa_\om h^{\rho\om} - \eta_{\mu\nu} 
\Box h
+ \pa_\mu \pa_\nu h)
\nonumber
\\
&&
+ \frac{2}{3} \left[f_2(\Box) - f_0(\Box) \right] \, \frac{1}{\Box} \, \pa_\mu \pa_\nu \pa_\rho \pa_\om 
h^{\rho\om}
= - 2 \ka  \, T_{\mu\nu},
\eeq 
where $T_{\mu\nu}$ is the energy-momentum tensor sourcing the field.
In the non-relativistic limit, for a static and spherically symmetric mass distribution $\rho(r)$ 
one has $T_{\mu\nu} = \rho\, \de_\mu^0 \, \de_\nu^0 \,$ and the metric
can be written in isotropic coordinates,
\beq
\label{m-New}
ds^2 \,=\, - (1+ 2 \ph) dt^2 + (1 - 2 \psi) (dx^2+dy^2+dz^2) \, ,
\eeq
where $\ph(r)$ and $\psi(r)$ are the Newtonian-limit potentials
and $r \,=\, \sqrt{x^2+y^2+z^2}\,$. These two independent potentials 
are obtained by solving the 00-component and the trace of the
equations of motion~\eqref{lieq},
\beq
\label{2-maz}
&&
2 f_2(\lap) \lap (\ph + \psi) - 2 f_0(\lap) \lap (\ph - 2 \psi ) = 3 \ka \rho
\,,
\\
&&
\label{eq1-maz}
2 f_0(\lap) \lap (\ph - 2 \psi) \,\,\,= - \ka  \rho
\, .
\eeq

Higher-derivative gravity models generally contain scalar and tensor massive excitations, and the potentials $\ph$ and $\psi$ depend on these masses. It is possible to separate the contribution of each spin sector by splitting the original potentials into auxiliary ones, $\chi_0$ and $\chi_2$, via~\cite{BreTib1}
\beq
\label{Chi-Ome-Def}
\ph = \frac{1}{3} (2 \chi_2 + \chi_0) \, , \qquad \psi = \frac{1}{3} (\chi_2 - \chi_0) \, .
\eeq
With these definitions eqs.~\eqref{2-maz} and~\eqref{eq1-maz} simplify to
\beq
\label{eqA}
&&
f_s (\lap) \lap \chi_s  =   \ka_s \, \rho \, ,
\eeq
where $\ka_0 = -\ka/2$, $\ka_2 = \ka$, and the functions $f_s$ (with $s=0,2$) on the \textit{l.h.s.} are precisely the ones which define the poles of the propagator~\eqref{prop}. It is clear that the potential $\chi_s$ only depends on the spin-$s$ massive modes of the theory, as claimed. It is also useful to notice that the particular case in which $f_0 = f_2$, which is equivalent to having $F_1 = -2F_2$, yields $\ph = \psi = \chi_2/2$ and in this sense only one equation in~\eqref{eqA} has to be solved.

Once the potentials $\chi_0$ and $\chi_2$ are known, it is possible to evaluate the curvature invariants associated to the metric~\eqref{m-New}.
It turns out that the finiteness of the potentials is not enough to guarantee a regular Newtonian limit, as there can still be curvature singularities. The curvature invariants are finite at $r=0$ if and only if $\chi'_0(0) = \chi'_2(0) = 0$ (see, e.g.,~\cite{BreTib1,Frolov:Poly}). In this spirit, a finite potential $\chi_s$ is said to be regular if $\chi^\prime_s(0)=0$. 
Note that if $\chi_0$ is regular, then $R$ is also regular. Similarly, the regularity of $\chi_2$ implies in the one of $C_{\mu\nu\alpha\beta}^2$ (here $C_{\mu\nu\alpha\beta}$ is the Weyl tensor), as it
depends only on  the tensor sector~\cite{BreTib1}. In order to have regular curvature invariants $R_{\mu\nu}^2$ and $R_{\mu\nu\alpha\beta}^2$ it is necessary that both $\chi_0$ and $\chi_2$ be regular.

\section{Effective smeared sources in polynomial HDG}
\label{Sec3}

Each one of the eqs.~\eqref{eqA} can be viewed as a Poisson equation with a modified source,
\beq
\label{EqEffSour}
\lap \chi_s =  \ka_s \, \varrho_s \, ,
\eeq
where the ``new'' sources $\varrho_s$ satisfy
\beq
\label{InvFs}
\rho(r) = f_s(\lap) \, \varrho_s (r).
\eeq
Particularizing the consideration for a delta source $\rho(\textbf{r}) = M \delta^{(3)}(\textbf{r})$, one gets
\beq
\label{rho_def}
\varrho_s(r) = \frac{M}{2 \pi^2 r} \int_{0}^{\infty} \frac{k \sin(kr)}{f_s(-k^2)} dk \ .
\eeq
Note that the non-constant term $f_s(-k^2)$ in the integrand induces a smearing of the original delta source.

It is important to notice that the definition of $\varrho_s$ through~\eqref{InvFs} depends on the shape of the original source~$\rho$ and involves the inversion of the operator $f_s(\lap)$, which in general is not direct. In the present work we restrict considerations to the Dirac delta source, associated to a point-like mass in rest, which admits a Fourier representation. Moreover, as mentioned before, here we assume that $f_s>0$ on the real line, $f_s(0)=1$ and that $f_s(z)$ (if not trivial) diverges at least linearly as $z \longrightarrow \infty$. The last hypothesis is certainly verified for all polynomial functions, while for non-polynomial ones it acts as a constraint on the type of non-locality of the theory. Under these assumptions the
Fourier kernel associated to the function $1/f_s(-k^2)$ is well-defined on the space of square-integrable functions and allows one to define the source $\varrho_s$ through its Fourier transform, a 
standard procedure in the field of local and non-local HDG (see, e.g.~\cite{Tseytlin95,Modesto12,Zhang14,Li15,Myung17,Buoninfante:2018b,Jens,Modesto-LWBH}). Still, we point out that for static solutions the original d'Alembert operator $\Box$ is substituted by the Laplacian $\lap$, which avoids all the issues related to the choice of the appropriate Green function of the inverse operator (see, e.g., the discussion on~\cite{Maggiore1} for the case of IR-modified theories, and on~\cite{Calcagni-Universe} for the case of non-perturbative solutions).

In what concerns locality, up to this point we did not make any \textit{a priori} restriction on the nature of the functions $f_s$ (or, equivalently, on $F_1$ and $F_2$ in the action).
Let us now assume that $f_s$ are polynomials; in other words, let us consider the case of polynomial HDG~\cite{AsoreyLopezShapiro}. Owed to the fundamental theorem of algebra, if $f_s$ has degree $\mathcal{N}_s$ and it is such that $f_s(0)=1$ (see eqs.~\eqref{f0} and~\eqref{f2}), then it can be written as
\beq \label{PolyFactor}
f_s(-k^2) = \prod_{i=1}^{N_s} 
\left( \frac{k^2 + m_{(s)i}^2 }{m_{(s)i}^2} \right) ^{n_{(s)i}}\, .
\eeq
Here we assume that the equation $f_s(-k^2)=0$ has $N_s$ roots $-m_{(s)i}^2$ (with $i=1,...,N_s$), each of them with multiplicity $n_{(s)i}$. It is clear that $\sum_i n_{(s)i} = \mathcal{N}_s$.
Accordingly, we can expand the term $1/f_s(-k^2)$ in~\eqref{rho_def} in partial fractions,
\beq
\label{partfrac}
\frac{1}{f_s(-k^2)} = \sum_{i=1}^{N_s} \sum_{j=1}^{n_{(s)i}} \frac{\alpha_{(s)i,j}}{(k^2 + m_{(s)i}^2)^j} \, ,
\eeq
where $\alpha_{(s)i,j}$ are coefficients which can be easily calculated, e.g., by means of Heaviside's residue method. It is possible, however, to obtain useful results without the need of explicitly writing down the expression for them, as we show in the next sections (see also~\cite{BreTib1}).

Inserting~\eqref{partfrac} into~\eqref{rho_def} one obtains the effective source for a polynomial HDG,
\beq \label{rho_gen}
\varrho_s(r) = \frac{M \sqrt{\pi}}{(2\pi)^2} \sum_{i=1}^{N_{s}} \sum_{j=1}^{n_{(s)i}} \frac{\alpha_{(s)i,j}}{(j-1)!} \left( \frac{r}{2m_{(s)i}} \right)^{j-\frac{3}{2}} K_{j-\frac{3}{2}}(m_{(s)i} r) \, ,
\eeq
where $K_\nu$ is the modified Bessel function of the second kind.
As we are considering a general polynomial model, the expression above holds for degenerate modes --- which are explicitly taken into account through the summation over $j \in \lbrace 1,...,n_{(s),i}\rbrace$ --- and also for complex modes. In what concerns the latter ones, their masses were chosen with positive real part so that $\varrho_s(r)$ decreases to zero for large distances (see~\cite{Newton-BLG,BreTib1,ABS-large} for further discussion).

From~\eqref{rho_gen} it follows that the presence of complex poles in the propagator yields oscillatory contributions to the effective source. For example, for the sixth-order gravity with conjugate poles $m_{(s)1}=\overline{m}_{(s)2}=a+ib$,
\beq
\label{rho_complex}
\varrho_s(r) = \frac{M (a^2 + b^2)^2 \sin(b r)}{8 \pi a b r} e^{-a r} \, .
\eeq
Such oscillations of the source have been noticed for particular theories in~\cite{Jens,Modesto-LWBH}; and before that it was shown to occur in the potentials of general theories with complex poles~\cite{Newton-BLG} (see also~\cite{Modesto16,ABS-large}).

\subsection{Regular effective sources}
\label{Sec3a}

The smearing of the $\delta$-source does not necessarily imply that the resultant effective source is free of singularities. However, we shall prove that for HDG the effective source is regular in most of the cases. Namely, we show that  $\varrho_s$ is finite if $\mathcal{N}_s > 1$, i.e., for models with more than four derivatives in the spin-$s$ sector of the action.

To this end, let us expand the general expression~\eqref{rho_gen} for the effective source around $r=0$. Taking into account the corresponding formulas for the modified Bessel functions~\cite{Grad} it follows
\beq
\label{rho_expans}
\varrho_{s}(r) = \frac{M}{4\pi r} A_{(s)1} \, + \, w_s \, + \, O(r) \, ,
\eeq
where $A_{(s)1} \equiv \sum_{i} \alpha_{(s)i,1}$ and $w_s$ is a constant.\footnote{It is possible to prove that $w_s \neq 0$; however, we postpone this demonstration to section~\ref{Sec5}.}
To show that the source is regular it then suffices to verify whether 
the coefficient $A_{(s)1}$, multiplying the divergent term $r^{-1}$, vanishes. We claim that
\beq \label{S1}
A_{(s)1} =  \left\{ 
\begin{array}{l l}
m_{(s)1}^2 \, ,  &  \text{if } \mathcal{N}_s = 1,\\
0 \, , &  \text{if } \mathcal{N}_s > 1.\\
\end{array} \right . \,
\eeq

As mentioned before, this claim can be demonstrated for the general case without the need of explicitly calculating all the quantities $\alpha_{(s)i,j}$, but only considering relations between them. In fact,
regrouping the \textit{r.h.s.} of eq.~\eqref{partfrac} into a single fraction one gets
\beq
\label{rhs}
\frac{\sum_{i} \sum_{j}\alpha_{(s)i,j} \left( k^2 + m_{(s)i}^2 \right)^{n_{(s)i}-j} \prod_{\ell \neq i} \left( k^2 + m_{(s)\ell}^2 \right)^{n_{(s)\ell}}  }{\prod_i \left( k^2 + m_{(s)i}^2 \right)^{n_{(s)i}}} \, .
\eeq
Then, by comparing the numerators of both sides of~\eqref{partfrac} and setting to zero the coefficients of the terms which depend on $k^2$, one obtains $\mathcal{N}_s$ relations between the quantities $\alpha_{(s)i,j}$ and the masses $m_{(s)i}$. In particular, for the term of highest order, proportional to $k^{2(\mathcal{N}_s-1)}$, it follows $\sum_{i} \alpha_{(s)i,1} = A_{(s)1} = 0$. Of course, the case $\mathcal{N}_s = 1$ is trivial as there is no term depending on $k$ in the numerator of~\eqref{rhs}, so $\alpha_{(s)i,1} = m_{(s)1}^2$. This proves~eq.~\eqref{S1}.

We say that the delta source is completely regularised if both $\varrho_0$ and $\varrho_2$ are finite. According to what was just proved, this occurs provided that $F_1$ and $F_2$ are polynomials of degree at least one (i.e., they are non-trivial polynomials) and\footnote{The condition $F_1\neq -3F_2$ ensures that $f_0$ is a non-trivial polynomial, see~\eqref{f0}.} $F_1 \neq -3F_2$.

As an explicit example, the effective sources for Stelle's fourth-order gravity,
\beq \label{RhoStelle}
\varrho_s(r) = \frac{M m_{(s)1}^2}{4\pi r} e^{-r m_{(s)1}} \, ,
\eeq
diverge as $r \rightarrow 0$. On the other hand, for a sixth-order gravity with a spin-$s$ pole of multiplicity two one gets
\beq \label{RhoDeg2}
\varrho_s(r) = \frac{M m_{(s)1}^3 }{8 \pi} e^{-r m_{(s)1}} \, ,
\eeq
which is regular. It is immediate to verify that~\eqref{rho_complex} is also regular.

\subsection{Effective mass functions}
\label{Sec3b}

Following the description of the higher derivatives' effects through Poisson equations with effective  sources $\varrho_s$, we shall define the mass function\footnote{Note that the definitions and the general discussion carried out in this subsection can be applied also to non-local HDG theories.}
\beq \label{MassFunctionF}
m_s(r) = 4\pi \int_0^r x^2 \varrho_s(x) dx
\eeq
as the total effective mass inside a sphere of radius $r$ centred in the origin, associated to the potential $\chi_s$. In the general case this function is no longer a constant, insomuch as the effective density functions are smeared and the total mass $M$ now fills the whole space. The function $m_s(r)$ can be written in terms of generalized hypergeometric functions by substituting~\eqref{rho_gen} in the equation above. We omit the result as the expression is not illuminating. Instead, it is more instructive to use the expansion~\eqref{rho_expans} in association to~\eqref{S1} to show that, near the origin,
\beq \label{m-force}
m_s(r) \sim  \left\{ 
\begin{array}{l l}
r^2 \, ,  &  \text{if } \mathcal{N}_s = 1,\\
r^3 \, , &  \text{if } \mathcal{N}_s > 1.\\
\end{array} \right . \,
\eeq

The effective mass $m_s$ above is related to the (modified) Newtonian force exerted on test particles. 
It is different from the massive quantity $\tilde{m}_s$ which appears in the expression of the potential
\beq
\label{m-pot}
\chi_s(r) = \frac{\ka_s \tilde{m}_s(r)}{r} \, .
\eeq
Indeed, with this \textit{Ansatz} the eq.~\eqref{EqEffSour} yields
\beq
\tilde{m}^{\prime\prime}_s(r) =  r  {\varrho}_s (r) \, ,
\eeq
whence
\beq
\label{m-pot-exp}
\tilde{m}_s(r) = \tilde{m}_s^\prime(0) \, r + \frac{r}{4\pi} \int_0^r \frac{m_s(x)}{x^2} dx     \, .
\eeq
The integration constant $\tilde{m}_s(0)=0$ is defined by recalling that the total mass $M$ is now delocalised. On the other hand, in view of~\eqref{m-force} and~\eqref{m-pot}, $\tilde{m}_s^\prime(0)$ is set by the requirement that $\chi_s \rightarrow 0$ for $r \rightarrow \infty$, and it gives the value of the potential $\chi_s$ at the origin. Therefore, one can say that the power series expansion of
$\tilde{m}_s$ around $r=0$ starts with the linear term, and the next term is of the same order of~\eqref{m-force}.
From the consideration above one concludes that the coefficient of the term $r^2$ in the series expansion of $\tilde{m}_s$ is non-zero if and only if $\mathcal{N}_s=1$.

\section{Finite and regular potentials in polynomial HDG}
\label{Sec4}

The smearing of the $\delta$-source is not a sufficient condition for the cancellation of the curvature singularities. As we show in this section, the smeared effective source in polynomial HDG yields a finite modified Newtonian potential; but to regularise the curvature invariants it is necessary to have regular effective sources.

In fact, the considerations of section~\ref{Sec3b} regarding the effective mass $\tilde{m}_s(r)$ shows that the small-$r$ behaviour of the potential $\chi_s$ is, up to a constant (see~\eqref{m-pot} and~\eqref{m-pot-exp}),
\beq \label{chis}
\chi_s(r) \sim  \left\{ 
\begin{array}{l l} 
r \, + \, \mathcal{O}(r^2) ,  &  \text{if } \mathcal{N}_s = 1,\\
r^2 \, + \, \mathcal{O}(r^3) \, , &  \text{if } \mathcal{N}_s > 1.\\
\end{array} \right . \,
\eeq
In both cases the potential is finite at $r=0$, with $\chi_s(0)=\ka_s\tilde{m}_s^\prime(0)$, but in the former there are  curvature singularities since $\chi_s^\prime(0)\neq 0$, as discussed in section~\ref{Sec2}.
Therefore, the complete regularization of the $\delta$-source coincides with the regularization of the curvature invariants.
This conclusion matches the considerations of~\cite{BreTib1}, where the general expression for $\chi_s$ was derived and it was shown that HDG models with at least six derivatives in both spin-0 and spin-2 sectors have regular curvature invariants in the non-relativistic limit. Moreover, the reasoning presented here offers an alternative demonstration of the results of~\cite{Newton-MNS,Newton-BLG} on the finiteness of the modified Newtonian potential in polynomial HDG.

In order to close this discussion on local HDG, it may be instructive to present some explicit examples. We start by the fourth-derivative model defined by polynomials $f_s(-k^2) = 1 + m_{(s)1}^{-2} k^2$. In this case the effective delta source is given by~\eqref{RhoStelle}, which yields the mass function
\beq \label{msStelle}
m_s(r) = M \left[ 1 - e^{-m_{(s)1} r} (1 + m_{(s)1} r)\right] = \frac{1}{2} M  m_{(s)1}^2 r^2 + \mathcal{O}(r^3) 
\eeq
and the auxiliary potential
\beq \label{ChiSStelle}
\chi_s(r) = -\frac{M \kappa_s }{4 \pi  r} (1 - e^{-m_{(s)1} r} ) = - \frac{M  m_{(s)1} \kappa_s}{4\pi} +  \frac{M  m_{(s)1}^2 \kappa_s}{8\pi}  r + \mathcal{O}(r^2).
\eeq
Both of them have the short-distance behaviour presented in eqs.~\eqref{m-force} and~\eqref{chis} with $\mathcal{N}_0=\mathcal{N}_2=1$ (complete fourth-derivative gravity).
Even though the potentials are finite, as $\chi_s^\prime(0)\neq 0$ the curvature invariants are not regular; indeed, 
 near the origin the Kretschmann scalar associated to the metric~\eqref{m-New} with auxiliary potentials~\eqref{ChiSStelle} behaves like 
\beq
R_{\mu\nu\al\be}^2 \approx 
\frac{8G^2 M^2 }{9 r^2}\left(m_{(0)1}^4 + m_{(0)1}^2 m_{(2)1}^2 + 7 m_{(2)1}^4\right)
.
\eeq
However, its divergence is less strong than in GR, or in the incomplete fourth-derivative model, for which $R_{\mu\nu\al\be}^2  \sim r^{-6}$.

The inclusion of a sixth-derivative in only one of the spin sectors cannot completely regularise the effective source (and the whole set of curvature invariants~\cite{BreTib1,BreTib3}). For example, in the incomplete sixth-order model with
\beq
f_0(-k^2) = 1 + 2 m_{(0)1}^{-2} k^2 + m_{(0)1}^{-4} k^4  , \qquad
f_2(-k^2) = 1 + m_{(2)1}^{-2} k^2,
\eeq
$m_2(r)$ and $\chi_2(r)$ are the same as in~\eqref{msStelle} and~\eqref{ChiSStelle} with $s=2$, but the scalar sector has a pole of order two. Using the effective source~\eqref{RhoDeg2} with $s=0$ it follows
\beq
m_0(r) &=& {M} \left[ -1 +  e^{ m_{(0)1} r} -  m_{(0)1} r \left( 1 + \frac{1}{2} m_{(0)1} r\right) \right]  e^{- m_{(0)1} r} = \frac{1}{6} M m_{(0)1}^3 r^3 + \mathcal{O}(r^4),
\nonumber
\\
\chi_0(r) &=&
-\frac{\kappa_0 M }{4 \pi  r} \left[
1 - \left( 
1
+ \frac{m_{(0)1} r}{2}
\right) e^{-m_{(0)1} r}
\right] = 
-\frac{\kappa_0 M m_{(0)1}  }{8 \pi } +  \frac{\ka_0 M m_{(0)1}^3}{48 \pi} r^2 + \mathcal{O}(r^3),
\nonumber
\eeq
which, again, agree with~\eqref{m-force} and~\eqref{chis} with $\mathcal{N}_0=2$.
The Kretschmann scalar behaves like
\beq
R_{\mu\nu\al\be}^2  \approx \frac{56 G^2 M^2 m_{(2)1}^4}{9 r^2}
\eeq
for $r \rightarrow 0$.
As the spin-2 sector is the one with only four derivatives, it is expected that the dominant divergent term near $r=0$ depends only on $m_{(2)1}$.

Only with the regularization of the source in both sectors the Kretschmann scalar becomes regular. Indeed, in terms of the previous example, if one includes a sixth-derivative in the spin-2 sector too by choosing, e.g.,
\beq
f_s(-k^2) = 1 + 2 m_{(s)1}^{-2} k^2 + m_{(s)1}^{-4} k^4 , \qquad s=0,2,
\eeq
then
\beq
R_{\mu\nu\al\be}^2 =
\frac{G^2M^2}{27} \left(5 m_{(0)1}^6+8 m_{(0)1}^3 m_{(2)1}^3 + 32 m_{(2)1}^6\right)
+ \mathcal{O}(r)
.
\eeq

\section{Effective smeared sources in non-local HDG}
\label{Sec5}

As it was mentioned in the Introduction, the description in terms of effective smeared sources is often used in the framework of ghost-free HDG. Therefore, instead of obtaining the expressions for the source $\varrho_s$ and the effective mass functions for particular theories (see, for example,~\cite{Modesto12,Zhang14,Li15,Myung17,Buoninfante:2018b,Jens}), in this section we focus on more general aspects which follow from the comparison with polynomial HDG.

Let us start by a particular family of models called ghost-free gravity of type $N$ ($\text{GF}_N$), which is defined by choosing the functions $f_0=f_2$ of the form~\cite{BGKM12,Head-On}
\beq
\label{f_GFN}
f_s(-k^2) = \exp{\left(\frac{k^2}{\mu^2}\right)^{N}} \,, \quad N \geqslant 1 ,
\eeq
where $N \in \mathbb{N}$ and $\mu$ is a massive parameter.
It was shown in~\cite{Head-On,EKM16}
that all these theories have a regular modified Newtonian potential when coupled to a $\delta$-source. In view of the discussion in the preceding sections, it is reasonable to think that the associated effective source is regular too.\footnote{In~\cite{Jens} the effective sources for some values of $N$ were explicitly calculated.
The solution for the smeared sources in a general $\text{GF}_N$ theory can be directly obtained by means eq.~\eqref{rho_def}, and it is a combination of generalized hypergeometric functions $_{0}F_{2(N-1)}$ multiplied by powers of $r$. We omit the explicit (cumbersome) expression as this section aims to more general results. The analogous complete solution for the potential was presented in~\cite{EKM16}.} In what follows we prove this statement by showing that the effective source for $\text{GF}_N$ theories can be obtained as the uniform limit of a sequence of sources of local HDG.

To this end, let us consider the polynomial gravity defined by the particular choice
\beq
f_s (-k^2) = f_{s,N_s,n} (-k^2) = \sum_{\ell=0}^n \frac{1}{\ell!} \left( \frac{k^{2}}{\mu_s^{2}} \right)^{N_s \ell} \, , 
\eeq
where $N_s \geqslant 1$ and $n \geqslant 2$ are natural numbers and $\mu_s$ is a massive parameter. According to eq.~\eqref{rho_def}, the corresponding effective source is given by
\beq
\label{rho_seque}
\varrho_{s,N_s,n}(r) = \frac{M}{2 \pi^2} \int_{0}^{\infty} g_{s,N_s,r,n}(k) dk \, , 
\eeq
with
\beq
\label{j}
g_{s,N_s,r,n}(k) = \frac{ k \sin(kr)}{r f_{s,N_s,n} (-k^2)} \, 
\eeq
for a fixed $r$. 
Since the sequence of (integrable) functions $\left\lbrace  g_{s,N_s,r,n} \right\rbrace_{n=2}^\infty$ is tight and converges uniformly to
\beq
\label{J}
G_{s,N_s,r}(k)= \frac{k \sin(kr)}{r \exp{(k^2 / \mu_s^{2})^{N_s}}} 
\eeq
on every compact $K \subset [0,+\infty)$, it follows that
\beq
\label{source_NLN}
\lim_{n\rightarrow\infty} \varrho_{s,N_s,n}(r) = \frac{M}{2 \pi^2} \int_{0}^{\infty} G_{s,N_s,r}(k) dk \equiv \varrho_{s,N_s}(r) \,
\eeq
for each $r \in (0,+\infty)$. Actually, as the sequence $\left\lbrace \varrho_{s,N_s,n} \right\rbrace_{n=2}^\infty$ is equicontinuous\footnote{This can be proved by noticing that the sequence $\left\lbrace \varrho_{s,N_s,n}^\prime \right\rbrace_n$ of the derivative of the sources is uniformly bounded on $[0,\infty)$.
} and uniformly bounded on $[0,+\infty)$, one can show that the limit $\varrho_{s,N_s,n}\rightarrow \varrho_{s,N_s}$ is uniform on $[0,+\infty)$. Thence, the limiting source $\varrho_{s,N_s}$ is also continuous and bounded. We stress that in the proof of the regularity of $\varrho_{s,N_s}$ we did not use its specific form, given by~\eqref{J} and~\eqref{source_NLN}.

In view of eq.~\eqref{rho_def}, it is immediate to verify that the source~\eqref{source_NLN} with $G_{s,N_s,r}$ given by~\eqref{J} is the one associated to the function~\eqref{f_GFN} with $N=N_s$ and $\mu=\mu_s$. Therefore, the effective source $\varrho_{N}(r)$ of the $\text{GF}_N$ models is the uniform limit of an equicontinuous sequence of (non-singular) sources associated to local HDG models; whence $\varrho_{N}(r)$ is regular too. Since the regularity of the source implies in the one of the potential,
the result of~\cite{Head-On,EKM16} on the regularity of $\text{GF}_N$ models is verified.

It is worthwhile to notice that the previous consideration applies directly to more 
general ghost-free theories defined by the functions
\beq
f_s(-k^2) = e^{P_s (-k^2)}
\,,
\eeq
where, here and in what follows, $P_s(z)$ is a real polynomial such that $P_s(0)=0$ and $P_s > 0$ for large $\vert z \vert$~\cite{EKM16}.

Nevertheless, this reasoning should be applied with caution to general ghost-free theories defined by an 
arbitrary entire function with enhanced UV-behaviour. On the one hand, even though the exponential of an entire function can always be written as a power series, which converges uniformly on compact sets, the sequence of functions analogous to~\eqref{j} might be not tight or integrable, making some sources of the sequence $\left\lbrace \varrho_{s,n} \right\rbrace_n$ ill-defined. It may be necessary to pass to a subsequence $\left\lbrace f_{s,n^\prime} \right\rbrace_{n^\prime}$ in order to have a well-defined sequence of sources. On the other hand, it is still possible that this sequence converges only pointwise on $(0,+\infty)$, owed to the violation of equicontinuity. Therefore, despite being a sequence of bounded functions defined on $[0,+\infty)$, the convergence at $r=0$ --- and, thus, the regularity of the limiting source --- is not guaranteed.

As an example, let us consider the case of
weakly non-local gravity theories~\cite{Modesto12,Kuzmin,Tomboulis}. These models are defined by form factors $f_s(-k^2)=e^{H_s(-k^2)}$, where $H_s(z)$ is an entire function such that $H_s(0) = 0$ and which behaves like $\ln[P_s(-k^2)]$ when $k \rightarrow \infty$. One simple choice is~\cite{Modesto12,Tomboulis}
\beq
\label{H}
H_s(-k^2) = \frac{1}{2} \left[ \ga + \Ga(0,P_s^2(-k^2)) \right] + \ln P_s(-k^2) \, ,
\eeq
where $\ga$ is Euler-Mascheroni constant and $\Ga(0,z)$ is the incomplete gamma function. The issue with these non-polynomial entire functions is the occurrence of an infinite number of sign changes in the coefficients of the power series. In fact,~\eqref{H} yields
\beq
f_s = 1 + \frac{P_s^2}{2} - \frac{P_s^6}{72} + \frac{P_s^8}{288} - \frac{P_s^{10}}{4800} - \frac{P_s^{12}}{8100} + \mathcal{O}(P_s^{14}) \, .
\eeq
Hence, if the series is truncated in a term with negative coefficient, the function defined by this partial sum will have a zero on the real line, possibly making the corresponding source ill-defined. Passing to the subsequence $\left\lbrace f_{s,n} \right\rbrace_n$ of partial sums truncated on the $n$-th term with positive coefficient, one gets a well-defined sequence of sources, all of them being regular due to the polynomial nature of $f_{s,n}$. This sequence $\left\lbrace \varrho_{s,n} \right\rbrace_n$ converges pointwise on $(0,+\infty)$, but it does not converge uniformly if the sequence of derivatives $\left\lbrace \varrho_{s,n}^\prime \right\rbrace_n$ is not uniformly bounded.

This is indeed what happens if $P_s(z)$ is a monomial of degree $N_s=1$. It is easy to see that in this case $\varrho_s(r)$ diverges in the origin, as for large $k$ one has $f_s(-k^2) \approx e^{\frac{\ga}{2}} k^{2}$. In this regime
\beq
\lim_{r \rightarrow 0} \frac{ k \sin(kr)}{r f_{s} (-k^2)} \approx e^{-\frac{\ga}{2}} ,
\eeq
which is not integrable on an unbounded interval.
Hence, $N_s=1$ implies that $\varrho_{s,n} \rightarrow \varrho_{s}$ pointwise on $(0,+\infty)$, but not uniformly, as
$\lim_{r \rightarrow 0} \varrho_{s}(r) = \infty$. The same arguments can be used to show that the limit $\varrho_{s,n} \rightarrow \varrho_{s}$ is uniform if the degree of $P_s(z)$ is $N_s \geqslant 2$, and in this case the source is regular (and the potential too). 

As a second example, one can consider the more general quasi-local form factor~\cite{Modesto12,Tomboulis}
\beq
\label{HKuzCap3}
H_s(-k^2) = \alpha_s \left[ \ga + \Ga(0,P_s(-k^2)) + \ln P_s(-k^2) \right]   ,
\eeq
where $P_s(z)$ is a polynomial of degree $N_s$.
Note that the previous example follows from the choice $\alpha_s = 1/2$ and the substitution $P_s \mapsto P_s^2$, while the form factor proposed by Kuz'min~\cite{Kuzmin} corresponds to $\al_s \in \mathbb{N}$ and $N_s = 1$. For large momentum it holds
\beq
\label{HKuzLimCap3}
\lim_{k \rightarrow \infty} \, e^{H_s(-k^2)} \, \approx \, e^{\alpha_s \ga} \,  k^{2 N_s \alpha_s}  \, ,
\eeq
whence, in this regime,
\beq
\lim_{r \rightarrow 0} \frac{ k \sin(kr)}{r f_{s} (-k^2)} \approx \frac{e^{-\alpha_s \ga}}{k^{2(N_s \alpha_s - 1)}} .
\eeq
It follows that the condition for having a regular source reads $N_s \alpha_s > 3/2$. In particular, for the Kuz'min form factor ($N_s=1$) the effective source is regular if $\al_s \geq 2$.

More generally, the regularity of the effective sources in  local and non-local HDG models follows from the UV behaviour of the functions $f_s$. This can be understood in light of some observations:
\begin{itemize}
\item[i.] $f_s$ does not change its sign because, as we restrict considerations to tachyon-free models, the equation $f_s(-k^2) = 0$ has no root for $k \in \mathbb{R}$.
\item[ii.] If for large arguments the function $f_s(z)$ grows faster than $z^{3/2}$, then   $G_{s,r}(k) = \frac{ k \sin(kr)}{r f_{s} (-k^2)}$ is integrable for any $r \in [0,\infty)$. So, $\lim_{r \rightarrow 0} \varrho_s(r) < \infty$, i.e., the effective source is regular.
\item[iii.] Since $r>0$ implies 
\beq
\vert G_{s,r} \vert = \frac{k \vert \sin(kr) \vert}{r f_s (-k^2)} < \frac{k^2}{f_s (-k^2)} = G_{s,0} \, ,
\eeq 
then
$
\int_0^\infty G_{s,r}(k) dk \leqslant \int_0^\infty \vert G_{s,r}(k)\vert dk < \int_0^\infty G_{s,0}(k) dk
$, which means that
$\varrho_s(r)$ achieves its maximum at $r=0$. In particular, $\varrho_s(0) \neq  0$.
\end{itemize}
Recall that a constant term in the power series expansion of $\varrho_s$ around $r=0$ gives $m_s(r) \sim r^3$ (see section~\ref{Sec3b}). Thus, it follows from the last observation above that for any theory with a regular potential, the leading non-constant contribution to $\chi_s$ for small distances is of order $r^2$. Particularizing for polynomial HDG, this yields the conclusion that $w_s \neq 0$ in~\eqref{rho_expans}.

\section{Conclusions}
\label{Sec6}

Local and non-local higher-derivative models have fruitful applications in the field of perturbative quantum gravity, as classical and quantum singularities which stem in GR can be smoothed out. This is ultimately related to the improved behaviour of the propagator in the UV regime and, therefore, it might be reasonable to think that regularity, at least at the linearised level, should be ubiquitous in these theories. In fact, in~\cite{BreTib1} we showed that all the polynomial gravity models with more than four derivatives in both scalar and tensor sectors are regular in the weak-field limit. In the present work we give an alternative proof of this result, based on the description of the higher-derivative's effects through an effective matter source. In this approach, increasing of the number of derivatives in the action can be viewed as implementing the regularization of the source: the singular point-like $\delta$-source in GR becomes a singular smeared source in fourth-derivative gravity, and it is regularised in theories with six and more derivatives. Furthermore, the considerations in terms of effective sources allow an almost straightforward extension to  non-local HDG theories.

Regularity properties of non-local ghost-free gravities have been intensively studied in recent years~\cite{Tseytlin95,Modesto12,Zhang14,Li15,Myung17,Jens,Buoninfante:2018b,BMS06,BGKM12,Frolov:Poly,Head-On,EKM16,Koshelev:2018hpt,ref10,FZN:Exp,Buoninfante:2018xiw,Ercan18}, and it is useful to notice that the key ingredient here seems to be not the non-locality of the interaction or the ghost-free condition, but (again) the behaviour of the propagator in the UV, as we give examples of regular models with ghosts and renormalizable ghost-free models with singularities. The use of effective sources makes this clear, at least in the weak-field limit.

Indeed, we showed that the effective sources for ghost-free theories defined by the exponential of a polynomial can be regarded as the uniform limit of a sequence of regular sources of polynomial theories, being, therefore, regular too. This also holds for quasi-local theories defined by form factors $f_s(-k^2)$ which behave like $k^n$, $n \geqslant 4$, in the UV. In this sense, the good regularity properties of these ghost-free gravities do not follow from non-locality or from the absence of ghosts; instead, they can be viewed as being inherited from the local polynomial theories. This is in agreement to the point of view that non-local gravities are the limiting theories when the degree of the polynomial goes to infinity, and therefore, it has an infinite number of (complex) poles hidden in the infinity~\cite{CountGhost}. Of course, this is different from what concerns  the avoidance of ghosts in the propagator~\cite{Modesto12,Krasnikov,Kuzmin,Tomboulis,BMS06,BGKM12} or extra propagating degrees of freedom~\cite{CMNa,CMNb,Calcagni-Universe}, which do require non-locality.

The results of the present work, together with~\cite{BreTib1}, motivates further investigations on the static spherically symmetric solutions in the full non-linear regime of polynomial and non-local gravities, and prospective relations between them. In fact, in~\cite{Stelle78,Stelle15PRL,Stelle15PRD} one can see that the singularity associated to the $\delta$-source in the linearised fourth-derivative gravity was preserved in the non-linear scenario. In what concerns local models with more than four derivatives, the numerical calculations presented in~\cite{Holdom} give evidence that the spherically symmetric solutions are regular. Similar discussions have been carried out for the
 non-local models~\cite{Buoninfante:2018xiw,Buoninfante:2018b,Koshelev:2018hpt,Koshelev:2017bxd,Buoninfante:2018xif,Calcagni:2018pro,Bambi:2016uda,Calcagni:2017sov}, which raises the interesting question of to which extent the relation between local and non-local models in the linear regime can be extended to the full non-linear one.

%%%%%%%%%%%%%%%%%%%%%%%%%%%%%%%%%%%%%%%%%%%%%%%%%%%%
%%%%%%%%%%%%%%%%%%%%%%%%%%%%%%%%%%%%%%%%%%%%%%%%%%%%
\section*{Acknowledgements}

B.L.G is grateful to CNPq--Brazil for supporting his Ph.D. project.
T.P.N. wishes to acknowledge CAPES for the support through the PNPD program.
B.L.G. is grateful to the Department of Physics of the 
Universidade Federal de Juiz de Fora for the kind hospitality during his visit, and to Y. Rodr\'{i}guez-L\'{o}pez for the useful discussions.

%%%%%%%%%%%%%%%%%%%%%%%%%%%%%%%%%%%%%%%%%%%%%%%%%%%%%%
%\section*{References}

\end{document}